\PassOptionsToPackage{numbers}{natbib}
\documentclass{article}

\usepackage[utf8]{inputenc}
 \usepackage[dblblindworkshop, final]{neurips_2025}
\usepackage{graphicx}
\usepackage{float}

\workshoptitle{GenProCC: 1st Workshop on Generative and Protective AI for Content Creation}

\usepackage[utf8]{inputenc} % allow utf-8 input
\usepackage[T1]{fontenc}    % use 8-bit T1 fonts
\usepackage{hyperref}       % hyperlinks
\usepackage{url}            % simple URL typesetting
\usepackage{booktabs}       % professional-quality tables
\usepackage{amsfonts}       % blackboard math symbols
\usepackage{nicefrac}       % compact symbols for 1/2, etc.
\usepackage{microtype}      % microtypography
\usepackage{xcolor}         % colors
\usepackage[numbers]{natbib}
\usepackage{listings}
\lstdefinelanguage{json}{
  basicstyle=\ttfamily\scriptsize,
  showstringspaces=false,
  breaklines=true,
  frame=single,
  numbers=none,
  literate=
   *{true}{{{\color{black}true}}}{4}
    {false}{{{\color{black}false}}}{5}
    {null}{{{\color{black}null}}}{4}
}
\lstdefinestyle{prompt}{
  basicstyle=\ttfamily\scriptsize,
  showstringspaces=false,
  breaklines=true,
  frame=single,
  numbers=none
}

\title{Node-Based Editing for Multimodal Generation of Text, Audio, Image, and Video}

\author{%
  Alexander Htet Kyaw \thanks{Work completed during internship at Microsoft} \\
  Massachusetts Institute of Technology\\
  Camrbidge, MA \\
  \texttt{alexkyaw@mit.edu} \\
  \And
  Lenin Ravindranath Sivalingam\\
  Microsoft Research \\
  Redmond, WA \\
  \texttt{lenin@microsoft.com} \\
}

\begin{document}

\maketitle

\begin{abstract}

We present a node-based storytelling system for multimodal content generation. The system represents stories as graphs of nodes that can be expanded, edited, and iteratively refined through direct user edits and natural-language prompts. Each node can integrate text, images, audio, and video, allowing creators to compose multimodal narratives. A task selection agent routes between specialized generative tasks that handle story generation, node structure reasoning, node diagram formatting, and context generation. The interface supports targeted editing of individual nodes, automatic branching for parallel storylines, and node-based iterative refinement. Our results demonstrate that node-based editing supports control over narrative structure and iterative generation of text, images, audio, and video. We report quantitative outcomes on automatic story outline generation and qualitative observations of editing workflows.  Finally, we discuss current limitations such as scalability to longer narratives and consistency across multiple nodes, and outline future work toward human-in-the-loop and user-centered creative AI tools.

\end{abstract}

\section{Introduction}

Recent progress in generative models makes it possible to easily create text, images, audio, and video \cite{gm_comprehensive_2020}. These capabilities open new possibilities in domains ranging from filmmaking to game design. Today, the dominant interaction paradigm for generative models based on prompts \cite{liu_pre-train_2023}. However, a single prompt typically doesn't fully capture the users intend \cite{wu_promptchainer_2022}. To address this, AI-driven content generation platforms must go beyond one-shot generation to support iterative, human-in-the-loop workflows where narrative structures can be guided, revised, and extended. 

% Creators need ways to specify high-level direction while having the flexibility to make targeted edits without restarting the entire generation. 
Storytelling presents a challenging opportunity for controllability due to the many ways narratives can be structured. Stories may unfold through sequential or branching events \cite{mirowski_co-writing_2023}. Existing AI-driven systems often struggle to satisfy these constraints, producing workflows that are linear, non-iterative, and fixed in structure, lacking the flexibility to support both high-level control and targeted edits.

Our approach represents stories as graphs of nodes, where each node corresponds to a scene or event that can be iteratively created, expanded, or edited through generative AI. This representation makes the narrative structure explicit and allows users to branch, reorder, or refine content.  The system combines two complementary interaction modes: natural language interaction with a large language model for high-level edits and node-based interaction for iterative refinements. Each node supports multi-modal generative AI output such as text, audio, images, and video. Our system contributes a node-based framework for AI-assisted storytelling by enabling (1) automated story node generation, (2) selective node-based media editing, and (3) iterative refinement through node branching. 

\section{Related Works}
\label{gen_inst}
Advances in model architectures and training are beginning to support editable video generation \cite{foo_ai-generated_2025}. For example, recent work explores ControlNet-style video conditioning with edge and depth maps \cite{zhang_adding_2023}, spatio-temporal encoders that fuse layout and motion cues using motion vectors \cite{ho_denoising_2020}, and trajectory-based controllers that decouple camera and object motion \cite{singer_make--video_2022}. Building on these advances, AI-driven content creation platforms are introducing interfaces that foreground user control, offering new ways for creators to structure and edit generative outputs. OpenAI’s Sora introduced a storyboard interface for sequential video generation, allowing users to break a script into discrete scenes and specify captions for each shot \cite{openai_sora_nodate}. Google’s Flow leverages Veo, Imagen, and Gemini to enable filmmakers to produce cinematic clips with scene-level continuity and camera control \cite{google_flow_nodate}. Similarly, Runway Gen-2 provides users with prompt-based video generation and timeline editing \cite{runwayml_research_nodate}.  These systems highlight a growing emphasis on controllability at the level of scenes and sequences, though they typically remain tied to linear editing metaphors and interfaces. 

Node-graph editors have been previously used in AI-driven creative workflows. ComfyUI, an open-source interface for diffusion models, lets users compose complex generation pipelines through modular graph representations \cite{xu_comfyui-copilot_2025}. In narrative design, visual scripting tools such as Twine have long demonstrated how stories can be authored and visualized as branching graphs \cite{carlon_educational_2022}. More recently, Geneva by Microsoft Research takes high level narrative descriptions and uses LLMs to generate graph based story  representations \cite{leandro_geneva_2024}. Previous work on AI-assisted storyboarding has transformed scripts into segments, with an emphasis on linear storyboard generation and image outputs, while lacking support for editable, controllable, multimedia workflows \cite{maharana_storydall-e_2022}. 

Our work builds on these directions by integrating a prompt based chat interface with a node-based story graph for multimodal  generation. Unlike linear interfaces, our system emphasizes iterative editing at the node level for generative content creation, enabling both targeted updates and node level branching, comparison, and  reordering.

\section{Methods}
\label{headings}

\begin{figure}[h]
    \centering
    \includegraphics[width=1\linewidth]{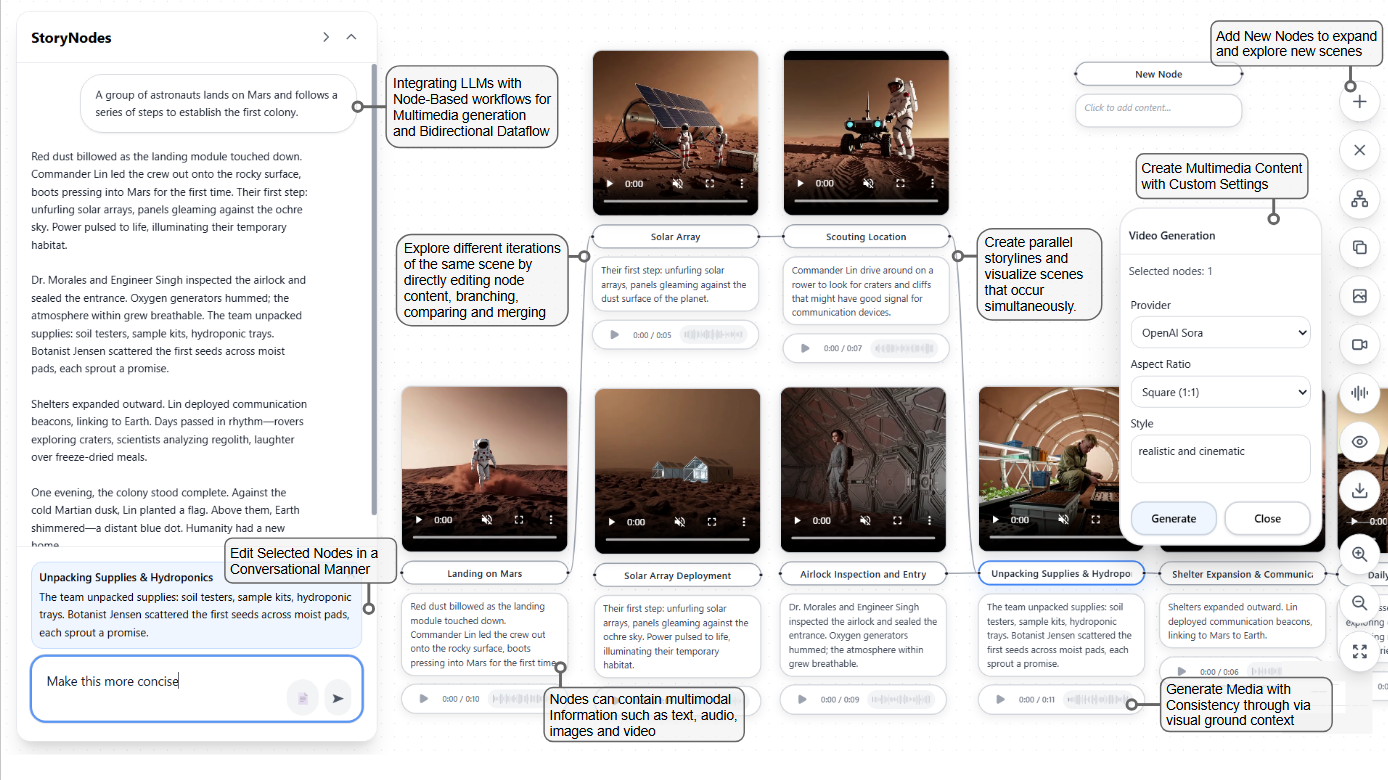}
    \caption{Conversational AI Interface integrated with a Node Based Multimedia Content Generation}
    \label{fig:interface}
\end{figure}

We present a storytelling platform that integrates a conversational interface with a node-based representation for multimedia generation. The system integrate natural language interaction with a structural graph view, enabling users to iteratively outline, compare, expand, and revise AI generated content (Fig. \ref{fig:interface}). This architecture allows both high-level story generation from prompts and specific updates to selected subset of nodes. 

% An task selection agent routes mechanism interprets user instructions and dispatches them to specialized generative components, including text generation, structural segmentation, diagram formatting, and multimedia synthesis.

\begin{figure}[h]
    \centering
    \includegraphics[width=1\linewidth]{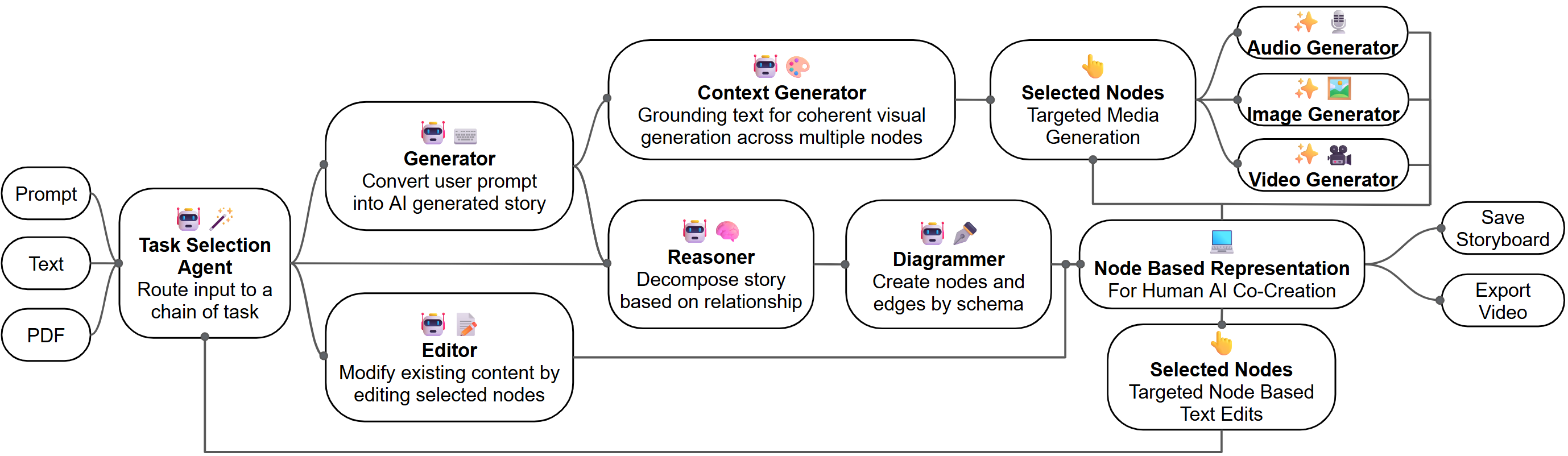}
    \caption{System Overview: From User Input, Task Selection Agent, Generator, Reasoner, Diagrammer, Node Graph Represenation, Editor, Context Generator to Multimedia Generation}
    \label{fig:system}
\end{figure}

\textbf{Large Language Models for Story Generation and Node-Based Diagram Generation. }A task selector agent acts as the orchestration layer of the system, interpreting user input, monitoring changes in the node diagram environment, routing requests to the appropriate generative task, and updating the diagram. Each of the tasks is driven by a large language model, specifically GPT-4.1 for its  reasoning and generation capabilities \cite{openai_overview_nodate}. When a user prompts for a story to be generated, the task selection agent routes the request to the Generator, which produces narrative text from the input (Fig \ref{fig:system}) The Reasoner then decomposes this text into a set of nodes, assigning edges that represent the narrative flow and relationships in the story. The Diagrammer formats the node list into a structured graph, enforcing strict JSON output with node titles, node segments, and edges (Appx \ref{json}). When the user wants to make targeted changes, the task selection agent routes to the Editor, which regenerates the selected nodes with user instructions while preserving the existing node structure. %Together, these tasks transform natural language prompts into interactive, editable story graphs rendered in the interface. 

% When users want to adjust layout without altering content, the agent routes to the Formatter, which uses GPT-4.1’s vision–language capabilities to interpret a diagram snapshot and reposition nodes for improved readability. 

\textbf{Node Based Representation for Multi-Media Generation}. The node structure acts as the interface between text generation and media generation. Stories in the system are represented as directed graphs, where each node contains a text segment and media assets such as audio, images, or video. The text segment serves as the prompt for generating the media linked to that node. For visual consistency, the system maintains a rolling story context that is passed to image and video generation tasks. % ensuring that characters, settings, and tone remain grounded to the same context.
Audio narration is produced by the GPT-4o's text-to-speech model, which converts each node’s text segment into audio narration (Fig. \ref{fig:sora}). Images are generated by GPT-Image-1 using each node’s text and contextual story details, while OpenAI’s Sora is used for video generation \cite{openai_overview_nodate} (Appx.~\ref{compute}). % which produces animated clips of each scene using the node description and broader story context as input. 
By structuring stories as graphs, users can select a node to create media for a single scene, regenerate assets for a subset of nodes, or preview the entire graph as an audiovisual sequence (Fig. \ref{fig:system}). Allowing users to maintain control over their AI-generated media, while preserving the graph structure.

\textbf{Graph Based Story Sequencing and Video Export.} Stories represented as graphs can be exported in multiple formats: as a compiled video clip, a visual storyboard, or a JSON graph. The full video export process handles the integration of audio, visuals, and subtitles (see Appx. \ref{export} Fig. \ref{fig:export}). Each path through the graph could corresponds to a possible narrative trajectory. Therefore, users have to option to select specific nodes to export or generate an export sequence based on the graph topology. % To maintain fidelity across exports, the backend validates node updates and clears stale assets whenever story content is regenerated. This ensures that exported outputs reflect the current graph and preserve narrative coherence while giving users control over how stories are transformed into shareable media artifacts.

\section{Experiments and Results}

\begin{figure}[H]
    \centering
    \includegraphics[width=1\linewidth]{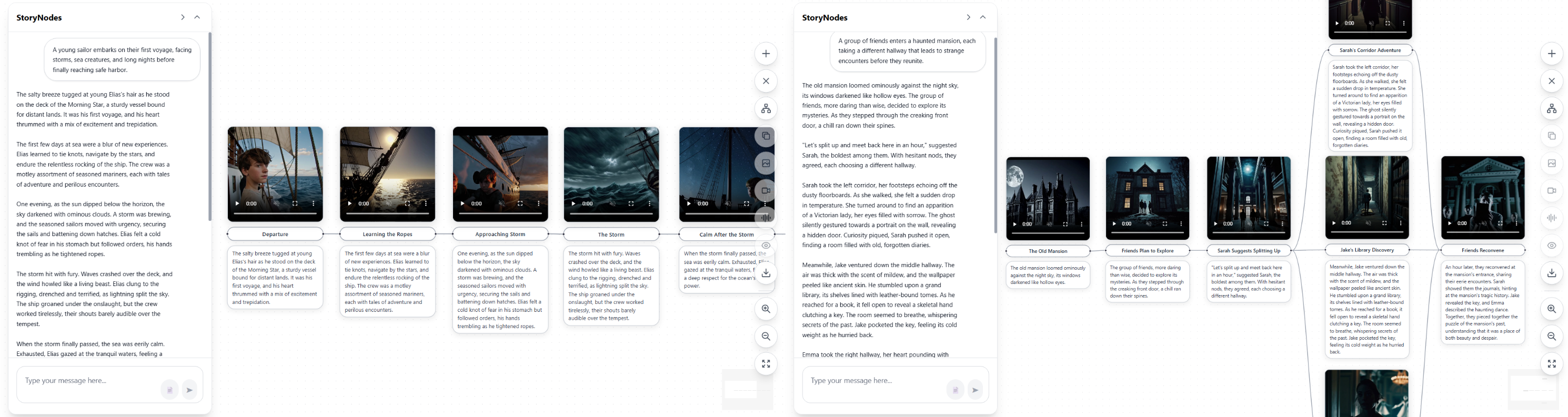}
    \caption{Generating a Sequential Storyline Generation versus a Branching Storyline}
    \label{fig:compare}
\end{figure}

\textbf{Automated story node generation.} To evaluate the system, we constrained the task to short stories represented as a graph of 8 to 12 nodes. Using a large language model (LLM), we generated 10 user prompts for single narrative stories and 10 user prompts for branching narrative stories. For single-narrative prompts, the system produced linear story graphs without unintended branching in 8 out of 10 trials. For branching prompts, the system generated parallel paths in 10 out of 10 trials (Fig. \ref{fig:compare}). These results indicate that the system can generate story graphs for both linear and branching narrative structures (see Appx. \ref{Evaluation} Fig. \ref{fig:evaluation1}, Fig.\ref{fig:evaluation2} for more details).

\begin{figure}[h]
    \centering
    \includegraphics[width=1\linewidth]{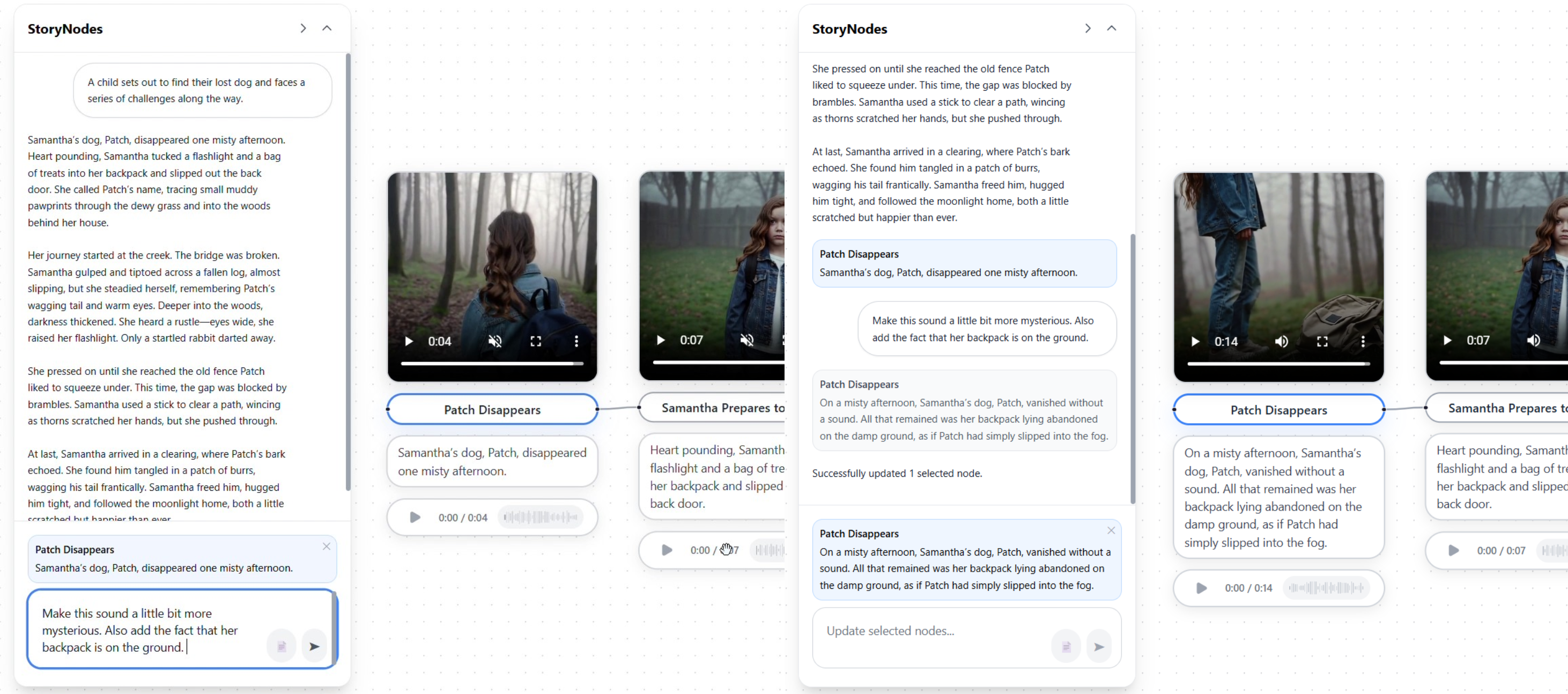}
    \caption{Using an LLM to edit selected nodes and make targeted changes to tone and story details}
    \label{fig:llmedit}
\end{figure}

\textbf{Selective node-based editing and content creation.} The interface supports manual edits directly in the node context and AI-driven edits. Manual editing was effective for targeted changes, such as altering specific objects, settings, or scene details, with edits immediately reflected in exported outputs. AI-assisted editing, allowed for higher-level modifications such as adjusting tone (e.g., "make this sound mysterious"), extending descriptions (e.g., "add the fact that her backpack is on the ground") or condensing narration (Fig. \ref{fig:llmedit}). We observed that manual editing was most effective for specific adjustments, while AI editing was more useful for structural or stylistic revisions. Once the node is regenerated or edited users can regenerate the media (Appendix \ref{Node-level}, Fig. \ref{fig:node}). In addition, the system supported global edits: users could select all nodes and request a rewrite that preserved the graph structure while modifying narrative details across the story (Appx \ref{Global}, Fig. \ref{fig:global}).

\textbf{Iterative refinement through branching and comparison.}
Unlike regular content, generative AI content can be regenerated and refined output through prompt changes. Our results demonstrated that a node-based representation can facilitate this AI-driven iterative exploration. Users could duplicate nodes or branches to create multiple versions of the same storyline, then compare them side by side. This allowed for experimentation with different stylistic directions while preserving the underlying structure. Such iteration was not possible in linear prompt-based systems, highlighting the advantage of the graph-based approach for exploring narratives with generative AI content.

%Our results demonstrate that a node-based, AI-driven interface enables more control and flexibility in multimodal content generation workflows

\section{Limitation and Conclusions}

%Our results demonstrate that the system can generate coherent story outlines from simple prompts, enable manual and AI-driven fine-grained editing, and support an iterative creative process with branching and revisions. This allow users to maintain narrative structure and make targeted modifications, addressing a common pain point of GenAI which can be uncontrollable or monolithic. 

A key limitation of this work is its reliance on text-based context grounding to maintain consistency across multiple generations. Future work could integrate image grounding for coherent media generation across multiple nodes or ground generated media grounded with real world data \cite{gupta_insights_2025}. Another limitation of the system is its ability to handle longer text and larger node graphs. Future work could explore hierarchical generation or subgraph-based approaches to preserve clarity and narrative coherence in longer context generation. We also plan to further conduct user studies with content creators to gather feedback on the usability of the interface and its impact on their creative process. By representing AI-generated content as author-able visual nodes, the system can lower barriers for non-technical users, preserve human agency, and increase access to collaborative human-AI creation for broader societal impact (Appx. \ref{Impact}). In conclusion, our work contributes a step toward AI-assisted creative content generation workflows that are controllable, editable, and iterative. 

Iterative narrative exploration allows creators to experiment with branching storylines, compare outcomes, and refine narratives through targeted regeneration. By duplicating nodes or branches, users can explore stylistic or thematic variations without disrupting the overall flow or regenerating the entire sequence. This process supports side-by-side comparisons of media and text, helping creators select the branch that best fits their vision. Branching iterations enable nondestructive content creation, which is especially valuable given the time and computational cost of generating video.

\begin{figure}[h]
    \centering
    \includegraphics[width=1\linewidth]{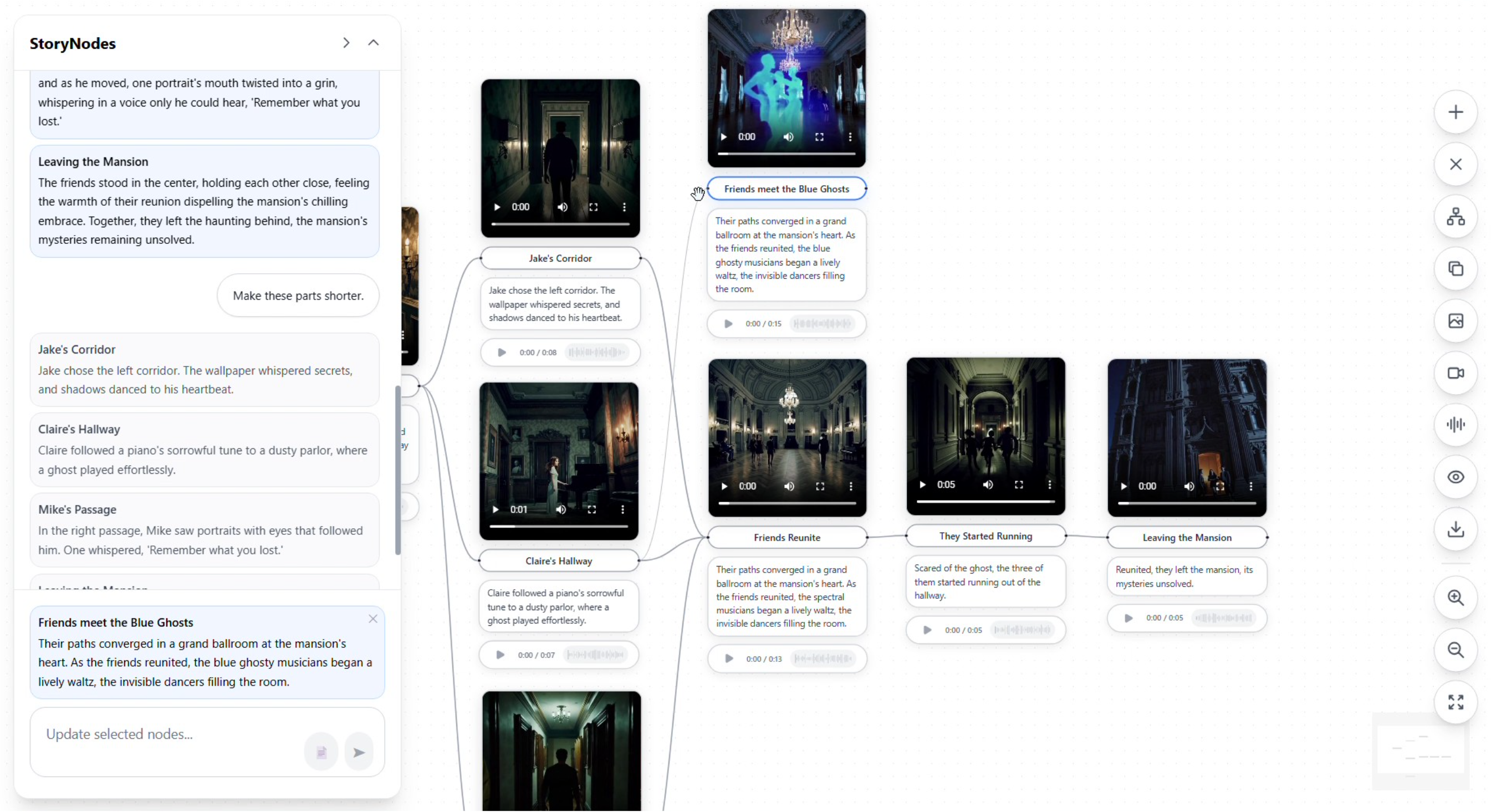}
    \caption{In this example, the “Friends meet the Blue Ghosts” node was regenerated with alternative descriptions, producing different video interpretations. The system displays both versions side by side, allowing the user to compare outcomes and select the preferred branch for continuing the narrative.}
    \label{fig:compare}
\end{figure}

\begin{figure}[H]
    \centering
    \includegraphics[width=1\linewidth]{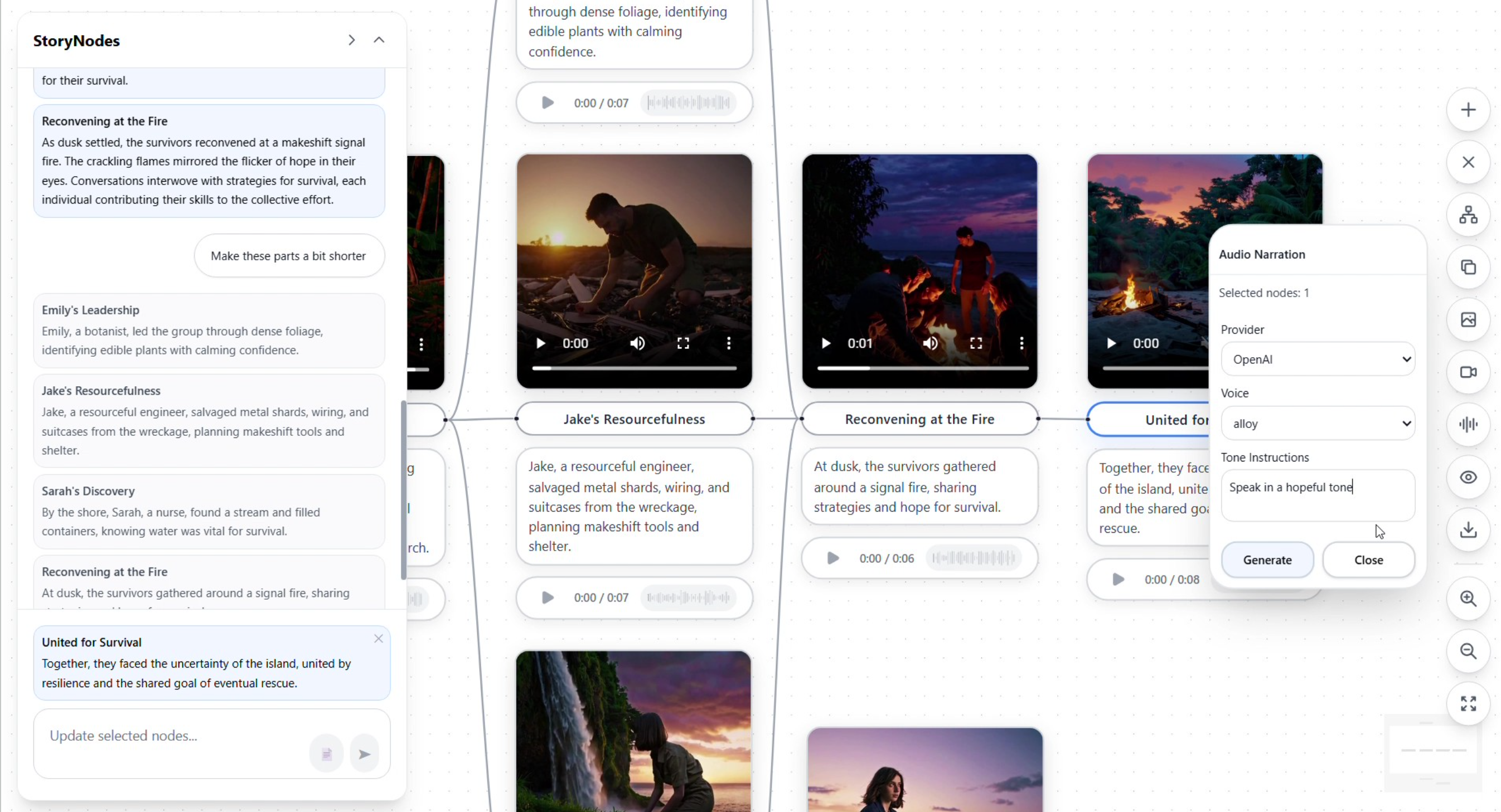}
    \caption{Media generation menu for audio narration. The author selects one or more nodes and configures provider, voice, and style instructions; generated clips attach to each card with duration and playback controls. In this case, the user instructed the AI model to speak in a hopeful tone.}
    \label{fig:sora}
\end{figure}

\begin{figure}[H]
    \centering
    \includegraphics[width=1\linewidth]{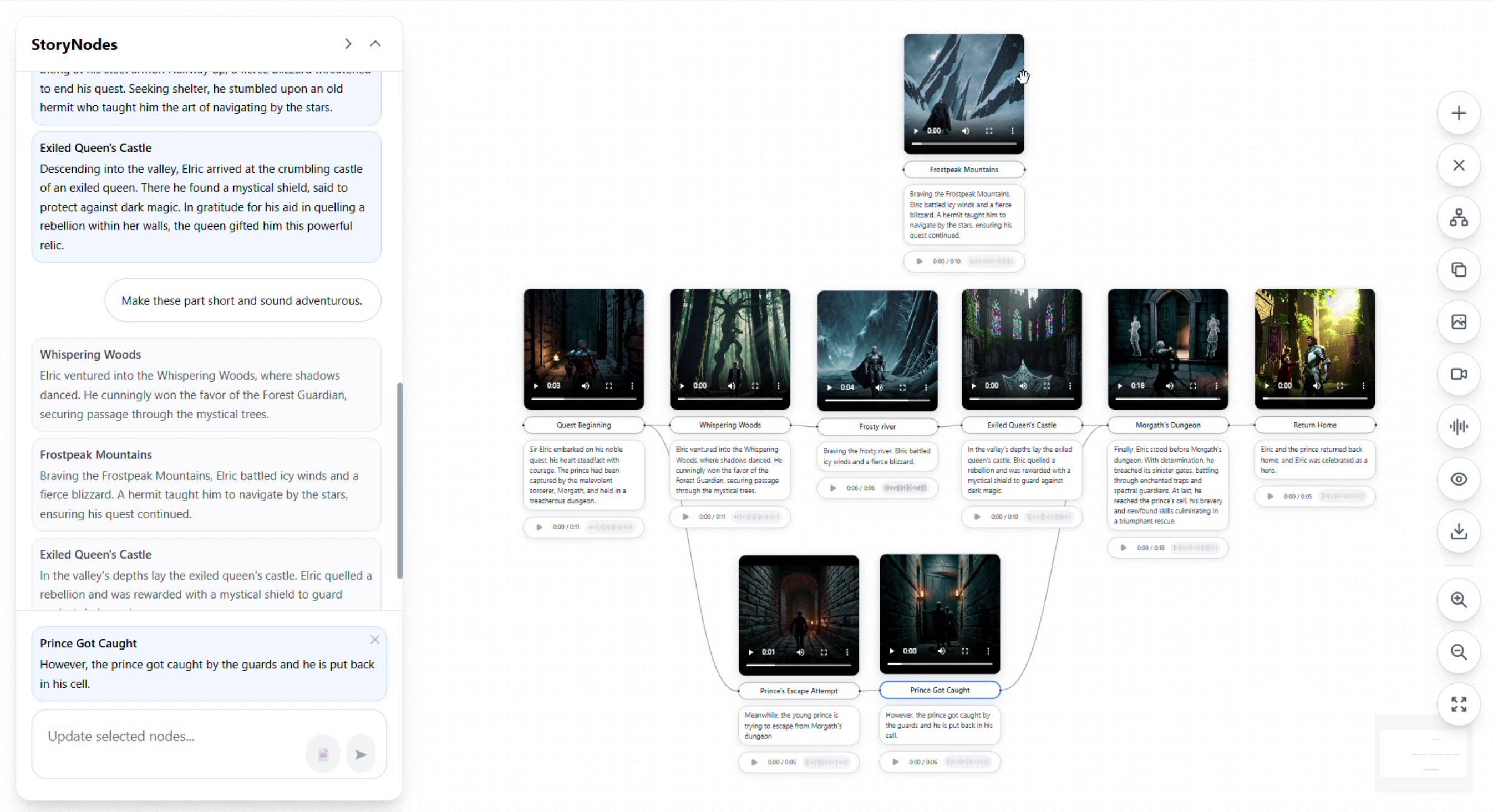}
    \caption{Example of branching parallel timelines within the story graph. This demonstrates how the system can represent events occurring at the same time in different narrative threads, allowing creators to explore and visualize parallel storylines.}
    \label{fig:branchcompare}
\end{figure}

\bibliographystyle{plainnat}
\bibliography{references}
%%%%%%%%%%%%%%%%%%%%%%%%%%%%%%%%%%%%%%%%%%%%%%%%%%%%%%%%%%%%

\appendix

\newpage

\section{Technical Appendices and Supplementary Material}

\subsection{Example of Structured Json Graph Output from Diagrammer LLM Task } \label{json}

\begin{lstlisting}[language=json,belowskip=0pt,aboveskip=4pt]
{"nodes":[
{"id":"1","data":{"label":"City of Lumina Plunged Into Darkness","segment":"The city of Lumina was known for its bright lights and bustling streets, but one fateful night, everything changed. Without warning, the entire city was plunged into darkness. Streetlights flickered out, neon signs went dim, and homes were cast into shadow."},"position":{"x":50,"y":50}},
{"id":"2","data":{"label":"Elena's Determination","segment":"In the upscale district of Crestwood, Elena, an emergency room nurse, decided she couldn't wait for the power to return. Her phone's battery was low, and she knew the hospital would need all hands on deck. Grabbing a flashlight, she ventured into the inky night, determined to make her way to the hospital on foot."},"position":{"x":350,"y":50}},
{"id":"3","data":{"label":"Marcus Offers Help","segment":"Across town in the industrial zone, Marcus, a mechanic, was about to close up his garage when the blackout hit. With a generator on hand, he decided to stay put, offering help to the stranded drivers who began to trickle in, searching for assistance in the darkened streets."},"position":{"x":350,"y":550}},
{"id":"4","data":{"label":"Friends' Night Adventure","segment":"In the heart of the city, at Central Park, a group of friends gathered for a night picnic found themselves enveloped in an unexpected adventure. Sarah, the group's unofficial leader, suggested they use the opportunity to explore the city's hidden corners under the cloak of darkness. They agreed, setting off with laughter and flashlights, their spirits undampened by the lack of light."},"position":{"x":350,"y":1050}},
{"id":"5","data":{"label":"Convergence at the Square","segment":"As the night wore on, Elena, Marcus, and Sarah;s group unknowingly moved towards a common destination-the city's main square. Elena trudged tirelessly, guided by the faint glow of her flashlight and sheer determination. Marcus, having helped as many as he could, decided to drive into the city to see if there was more he could do. Sarah's group, driven by youthful curiosity, meandered through alleyways and side streets."},"position":{"x":650,"y":300}},
{"id":"6","data":{"label":"Gathering at the Square","segment":"Their paths converged at the square, a place usually bursting with life but now eerily silent. Here, the community gathered, drawn by a mysterious glow emanating from a lone solar-powered art installation standing defiantly in the darkness. People shared stories, comfort, and resources, as the city, despite its fractured state, found unity in the shared experience."},"position":{"x":950,"y":300}},
{"id":"7","data":{"label":"Lights Return and Lasting Connections","segment":"As dawn approached, the lights flickered back to life, but the connections forged in the shadows lingered, leaving Lumina brighter than before."},"position":{"x":1250,"y":300}}
],
"edges":[
{"id":"e1-2","source":"1","target":"2"},
{"id":"e1-3","source":"1","target":"3"},
{"id":"e1-4","source":"1","target":"4"},
{"id":"e2-6","source":"2","target":"6"},
{"id":"e3-6","source":"3","target":"6"},
{"id":"e4-6","source":"4","target":"6"},
{"id":"e5-6","source":"5","target":"6"},
{"id":"e6-7","source":"6","target":"7"}]}
\end{lstlisting}

\subsection{Evaluation of Story Graph Generation}\label{Evaluation}

For automated story node generation, we evaluated 10 prompts for linear narratives and 10 prompts for branching narratives. The system produced correct linear graphs in 8/10 cases (80\%, 95\% CI [44\%--97\%]) and branching graphs in 10/10 cases (100\%, 95\% CI [69\%--100\%]). These binomial confidence intervals were calculated to reflect uncertainty given the small sample sizes, providing a measure of statistical significance for the reported success rates.

% --- Meta-prompts used to generate evaluation prompts ---
Below are the meta-prompts used to generate example prompts for the experiments to test branching and sequential narratives.

\begin{lstlisting}[style=prompt]
Generate 10 user prompts for a short story that can be represented as a branching narrative with parallel events. There should be around 8-12 events. The prompt should be around 1 to 3 sentences long. Return only the prompts, nothing else.

Generate 10 user prompts for a short story that can be represented as a linear sequence of events. There should be around 8-12 events in total. The prompt should be around 1 to 3 sentences long. Return only the prompts, nothing else.
\end{lstlisting}

\begin{table}[H]
\caption{Branching narrative prompts used in evaluation}
\label{tab:branching-prompts}
\centering
\begin{tabular}{@{}p{0.04\linewidth} p{0.86\linewidth} p{0.08\linewidth}@{}}
\toprule
\# & Branching narrative prompt & Result \\
\midrule
1 & A group of friends enters a haunted mansion, each taking a different hallway that leads to strange encounters before they reunite. & Pass \\
2 & A colony ship lands on an alien world, where different crew members explore separate regions that reveal conflicting discoveries. & Pass \\
3 & A royal court faces a crisis: the king seeks peace, the queen demands war, and advisors pursue secret plots that intertwine. & Pass \\
4 & A city is struck by a mysterious blackout, forcing residents across different neighborhoods to make choices that eventually converge. & Pass \\
5 & A team of treasure hunters splits up inside a vast cave system, each path filled with traps and clues pointing to the same artifact. & Pass \\
6 & A rebellion begins in a futuristic city, where different factions take divergent actions that may ultimately decide the same fate. & Pass \\
7 & A group of scientists investigates a spreading anomaly, with each researcher following a separate theory until their findings intersect. & Pass \\
8 & A traveling circus arrives in a new town, and performers’ separate adventures—on stage, in the streets, and in secret—eventually collide. & Pass \\
9 & A family separated during a natural disaster each struggles to survive in different locations, working toward reunion. & Pass \\
10 & A medieval village faces an approaching army, with villagers choosing to fortify defenses, hide in the forest, or negotiate, all leading to a shared resolution. & Pass \\
\bottomrule
\end{tabular}
\end{table}

\begin{table}[H]
\caption{Linear narrative prompts used in evaluation}
\label{tab:linear-prompts}
\centering
\begin{tabular}{@{}p{0.04\linewidth} p{0.86\linewidth} p{0.08\linewidth}@{}}
\toprule
\# & Linear narrative prompt & Result \\
\midrule
1 & A child sets out to find their lost dog and faces a series of challenges along the way. & Pass \\
2 & An archaeologist explores an ancient tomb, uncovering traps, puzzles, and a final treasure. & Pass \\
3 & A knight embarks on a quest to rescue a captured friend, passing through forests, mountains, and dungeons. & Pass \\
4 & A group of astronauts lands on Mars and follows a series of steps to establish the first colony. & Fail \\
5 & A chef attempts to prepare a complex dish, encountering difficulties but completing it step by step. & Pass \\
6 & A musician travels from town to town, slowly building recognition until reaching a grand concert. & Pass \\
7 & A fisherman battles a storm at sea, struggling with wind, waves, and exhaustion before making it back to shore. & Pass \\
8 & A teacher prepares their students for an important exam, overcoming obstacles in study sessions until the final test. & Fail \\
9 & A young inventor builds a flying machine, refining it through a series of trials until it finally succeeds. & Pass \\
10 & A messenger must deliver an important letter across dangerous terrain, encountering challenges one after another until the mission is complete. & Pass \\
\bottomrule
\end{tabular}
\end{table}

\begin{figure}[h]
    \centering
    \includegraphics[width=1\linewidth]{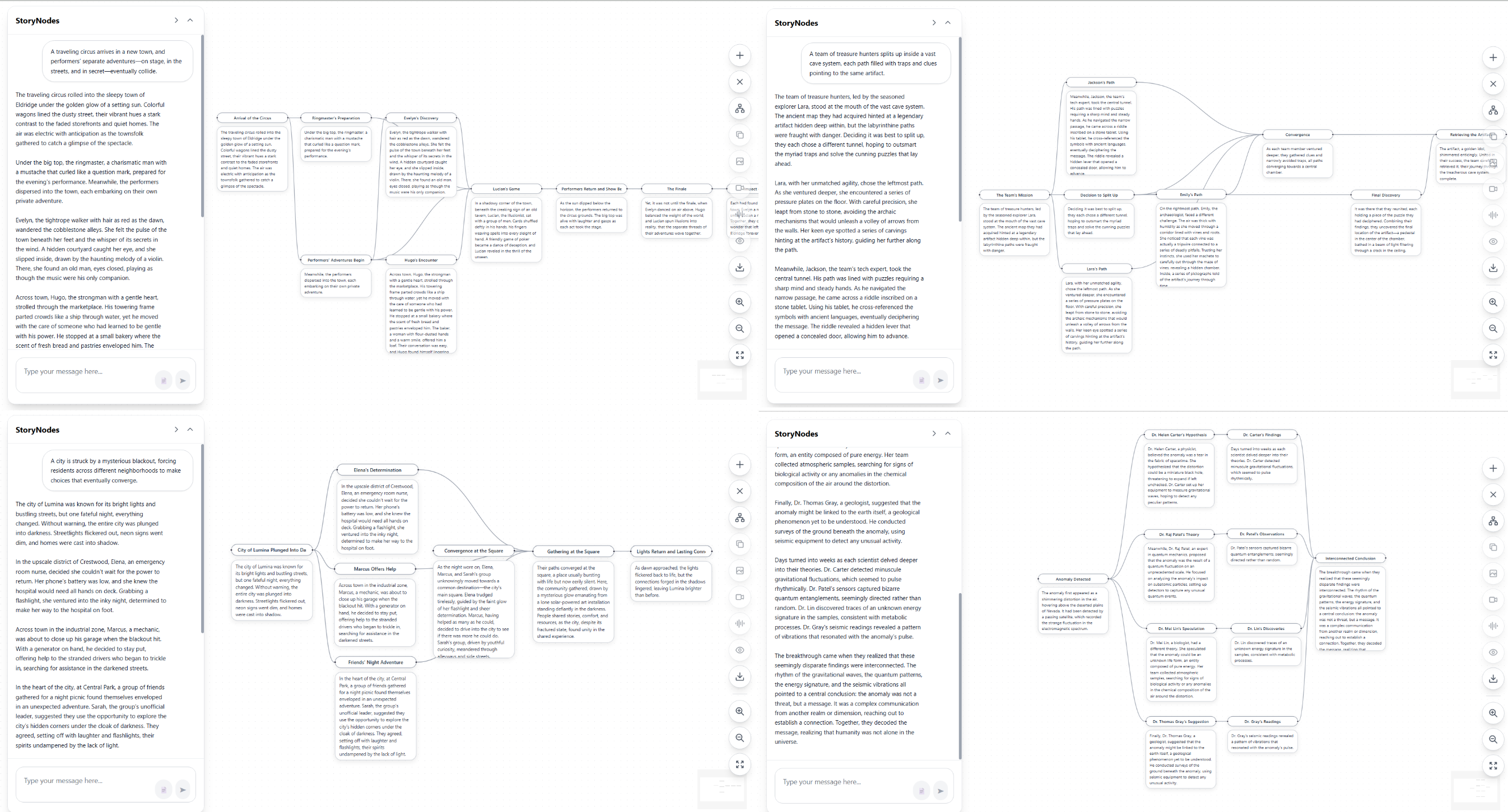}
    \caption{Example of Branching Narratives generated by the system}
    \label{fig:evaluation1}
\end{figure}

\begin{figure}[h]
    \centering
    \includegraphics[width=1\linewidth]{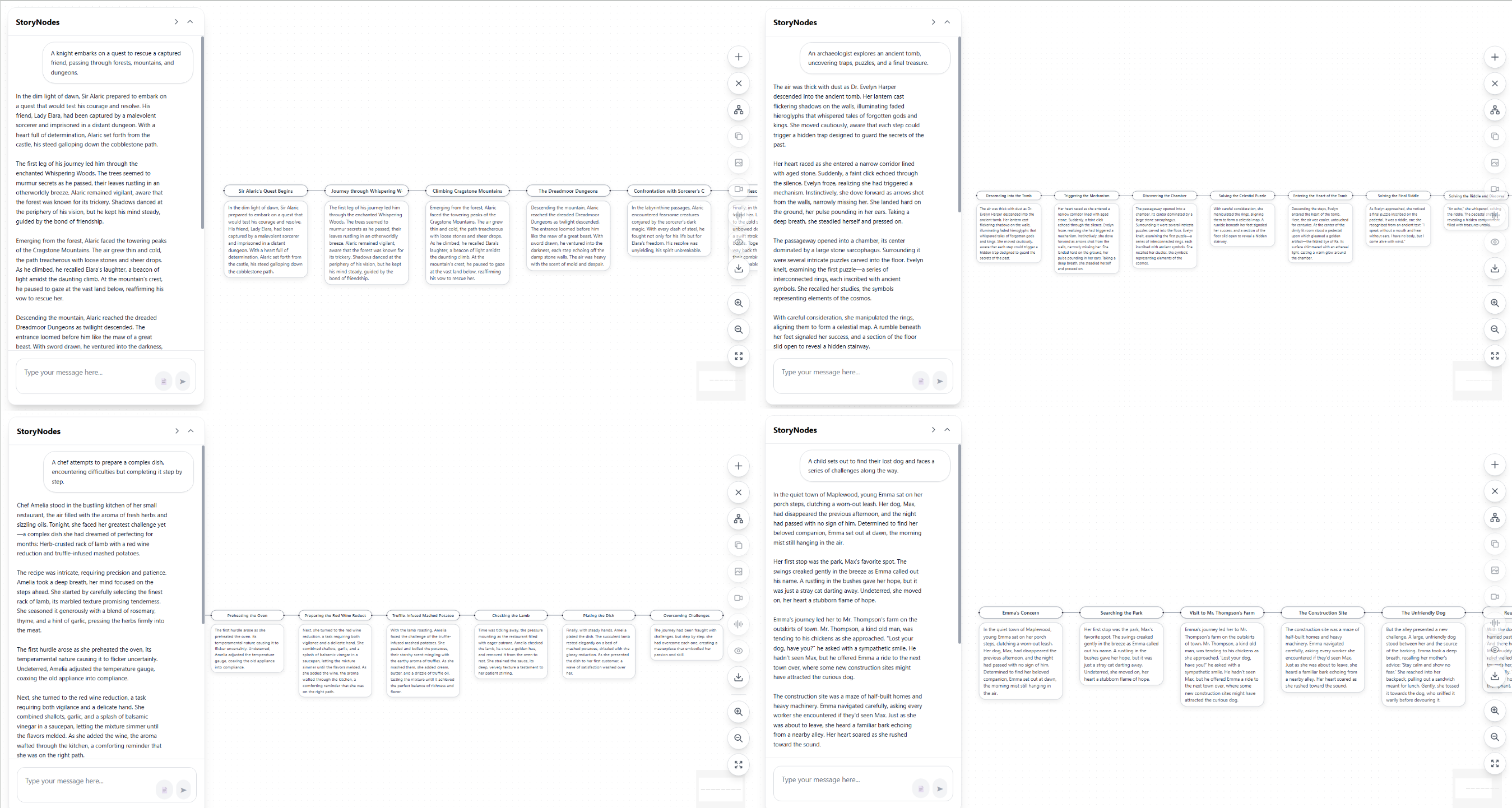}
    \caption{Example of Single Sequence Narratives generated by the system}
    \label{fig:evaluation2}
\end{figure}

\begin{table}[H]
\centering
\begin{tabular}{lccc}
\toprule
\textbf{Narrative Type} & \textbf{Correct / Total} & \textbf{Success Rate} & \textbf{95\% CI} \\
\midrule
Linear    & 8 / 10  & 80\%  & [0.44, 0.97] \\
Branching & 10 / 10 & 100\% & [0.69, 1.00] \\
\bottomrule
\end{tabular}
\caption{Evaluation of story node generation across 10 prompts for each narrative type.}
\label{tab:stats}
\end{table}

\newpage    

\subsection{Media Generation Menu and User Customization}\label{sora}

The media generation menu lets users create audio, images, and video at the node level. Users select any subset of nodes, set options such as voice and brief style instructions, and launch generation. Requests are queued, progress is shown inline on each node, and the resulting assets are written back to the selected nodes without altering their text. The system carries forward a rolling story context to promote consistency across nodes. Users can rerun the menu to regenerate only the selected nodes or batch apply new settings to a branch before exporting a preview or a compiled video.

\begin{figure}[h]
    \centering
    \includegraphics[width=1\linewidth]{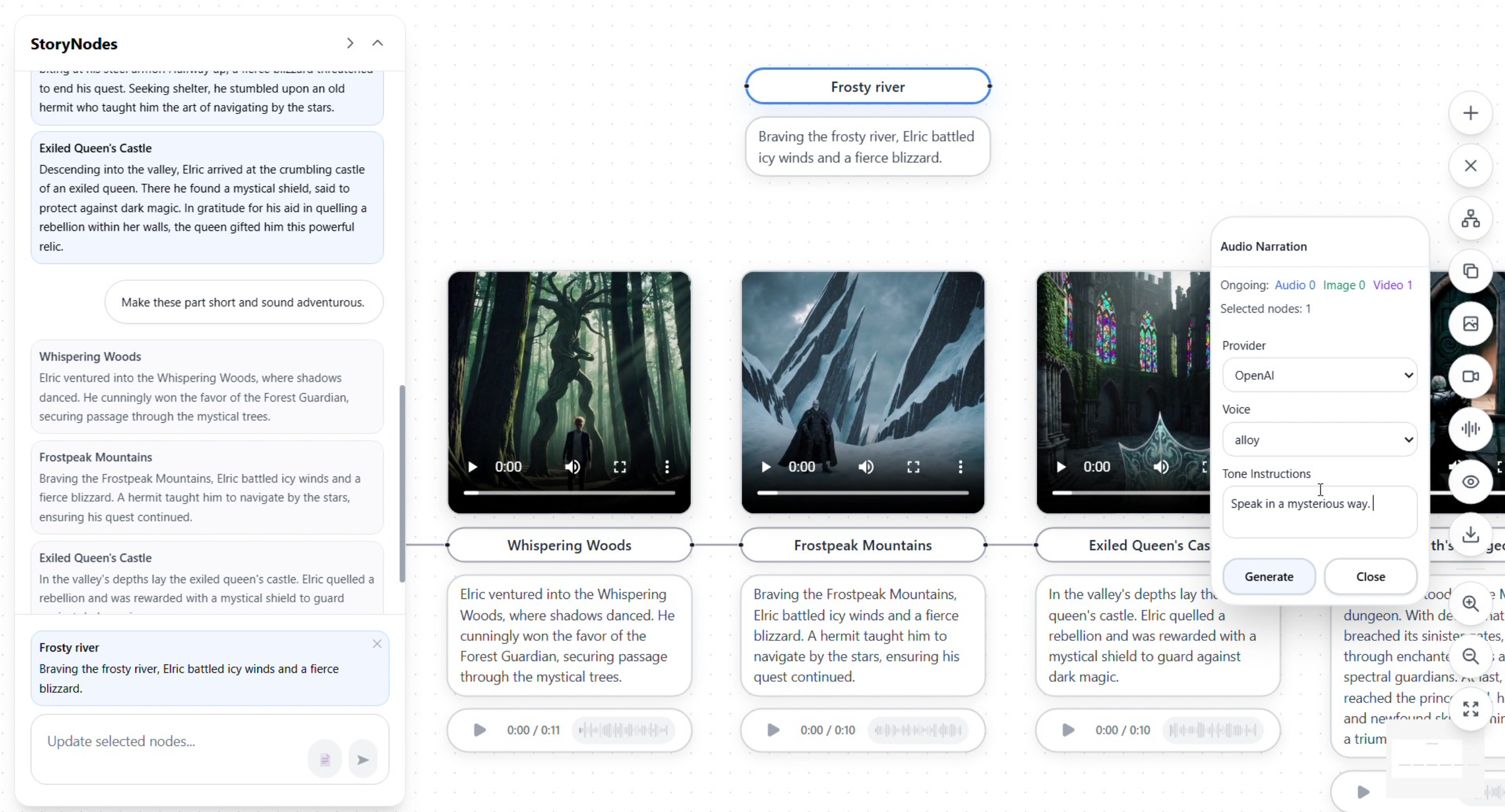}
    \caption{Media generation menu for audio narration. The author selects one or more nodes and configures provider, voice, and style instructions; generated clips attach to each card with duration and playback controls. In this case, the user instructed the AI model to speak in a mysterious way.}
    \label{fig:sora}
\end{figure}

\subsection{Node-Level Editing and Media Regeneration}\label{Node-level}

\begin{figure}[H]
    \centering
    \includegraphics[width=1\linewidth]{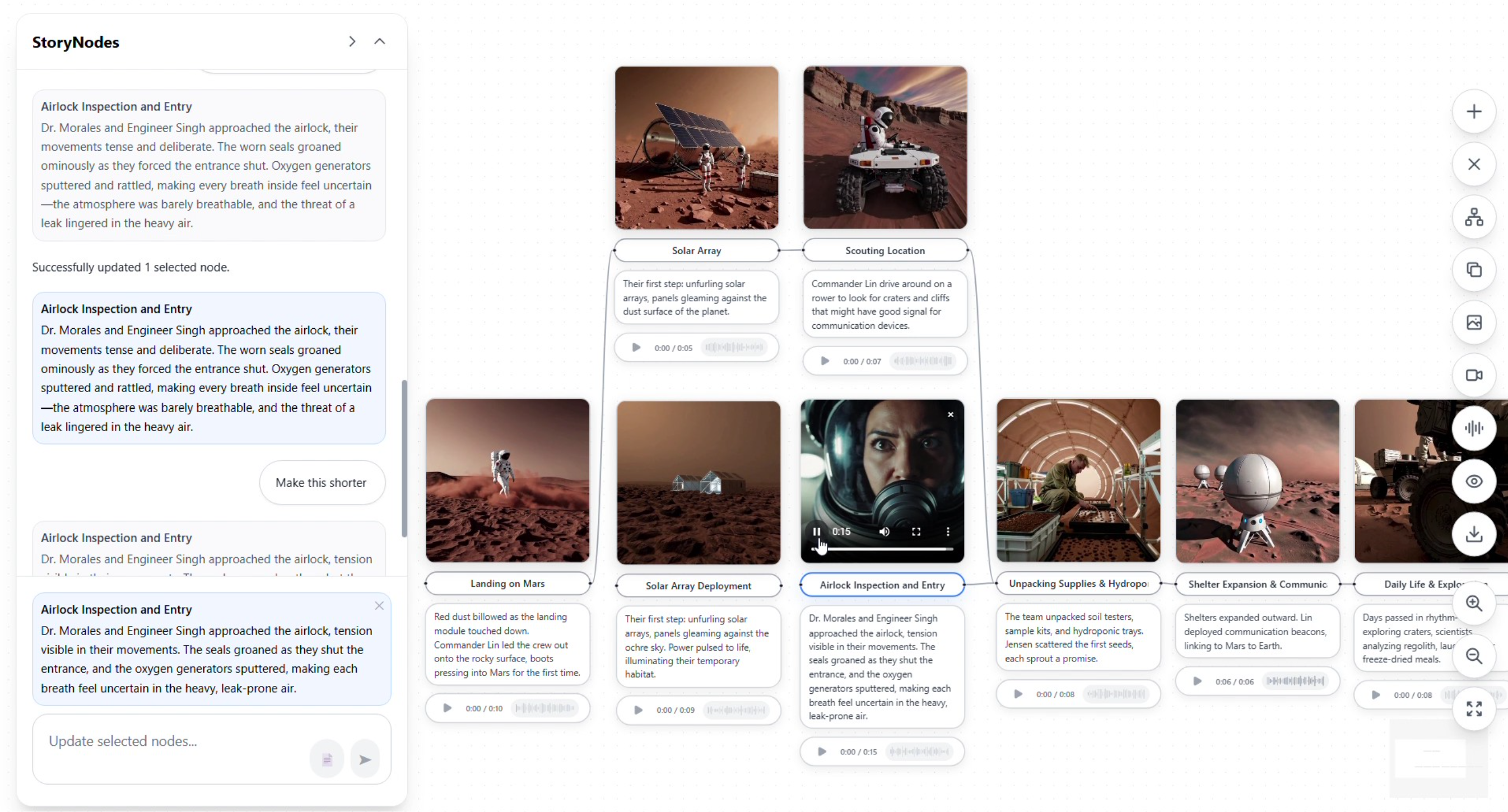}
    \caption{The Airlock Inspection and Entry node describes Dr. Morales and Engineer Singh approaching the airlock with visible tension}
    \label{fig:node}
\end{figure}

Node-level editing allows targeted revisions to specific parts of a story without regenerating the entire narrative. In the top figure, the Airlock Inspection and Entry node describes Dr. Morales and Engineer Singh approaching the airlock with visible tension. The associated media reflects this. In the bottom figure, the same node was edited to add details about the entrance. This textual change directly influenced the regenerated media, producing imagery with the people walking towards the entrance door. By comparing the two figures, we see how selective editing of a single node propagates into meaningful differences in the generated media while the rest of the story graph remains unchanged.

\begin{figure}[h]
    \centering
    \includegraphics[width=1\linewidth]{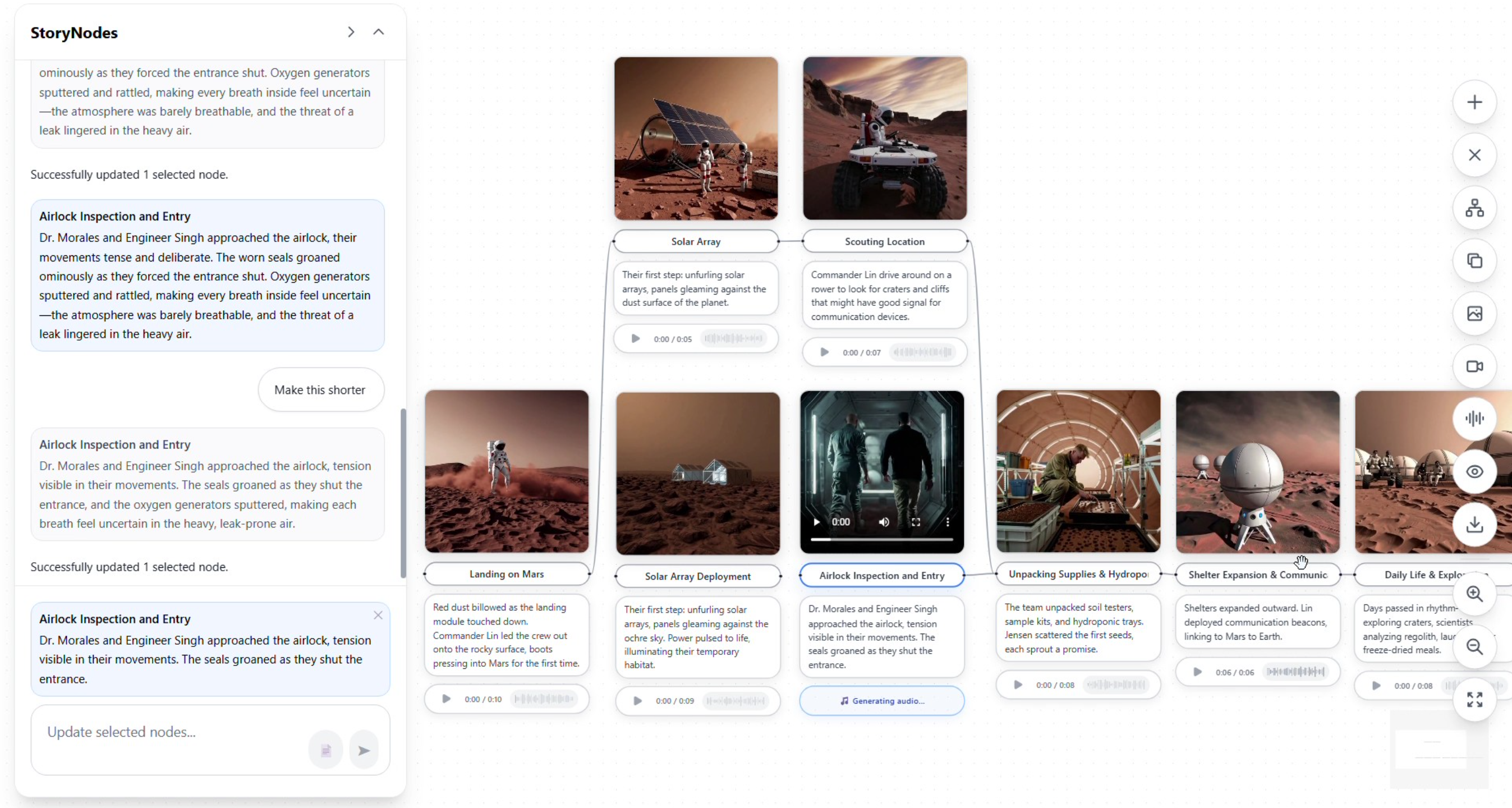}
    \caption{The Airlock Inspection and Entry node was edited to include details about the entrance, and the regenerated media reflects this change by showing the crew walking directly toward the airlock door.}
    \label{fig:node}
\end{figure}

\subsection{Node Based Global Graph Level Editing and Media Regeneration}\label{Global}

Graph-level editing enables simultaneous modification of multiple nodes, allowing creators to shift tone or style across entire story branches without altering the underlying structure. This global update not only streamlined the text but also influenced the associated media regeneration, producing visuals and audio. The overall graph structure remains unchanged, ensuring that edits adjust content while preserving narrative coherence. By operating at the graph level, the system provides a powerful mechanism for large-scale yet coherent adjustments, preserving narrative flow while adapting the creative direction across all branches.

\begin{figure}[H]
    \centering
    \includegraphics[width=1\linewidth]{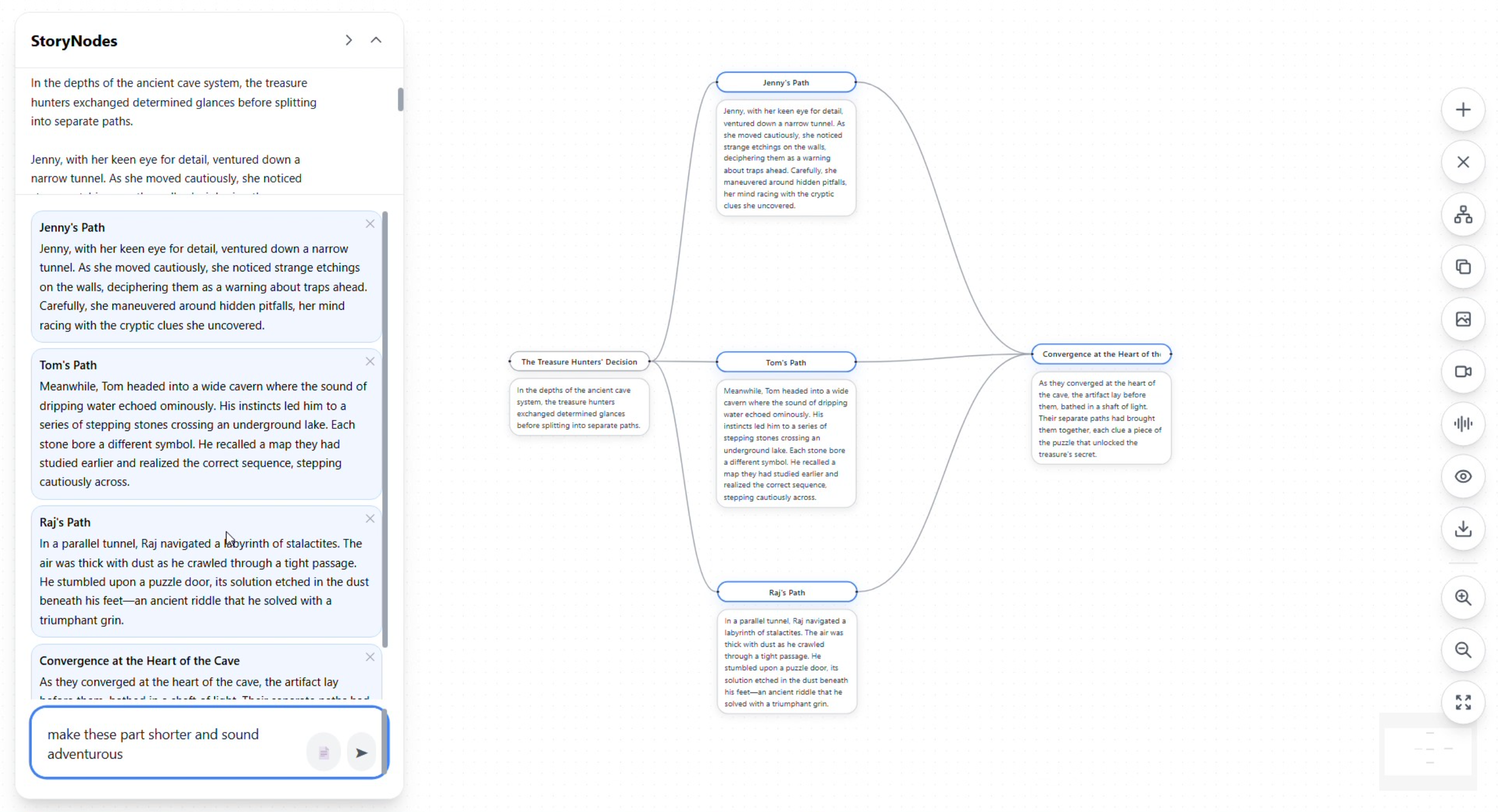}
    \caption{Initial version of the branching story graph, where each path (Jenny’s Path, Tom’s Path, Raj’s Path) is described in detail. The editing prompt “make these parts shorter and sound adventurous” is issued to update the narrative.}
    \label{fig:global}
\end{figure}

\begin{figure}[h]
    \centering
    \includegraphics[width=1\linewidth]{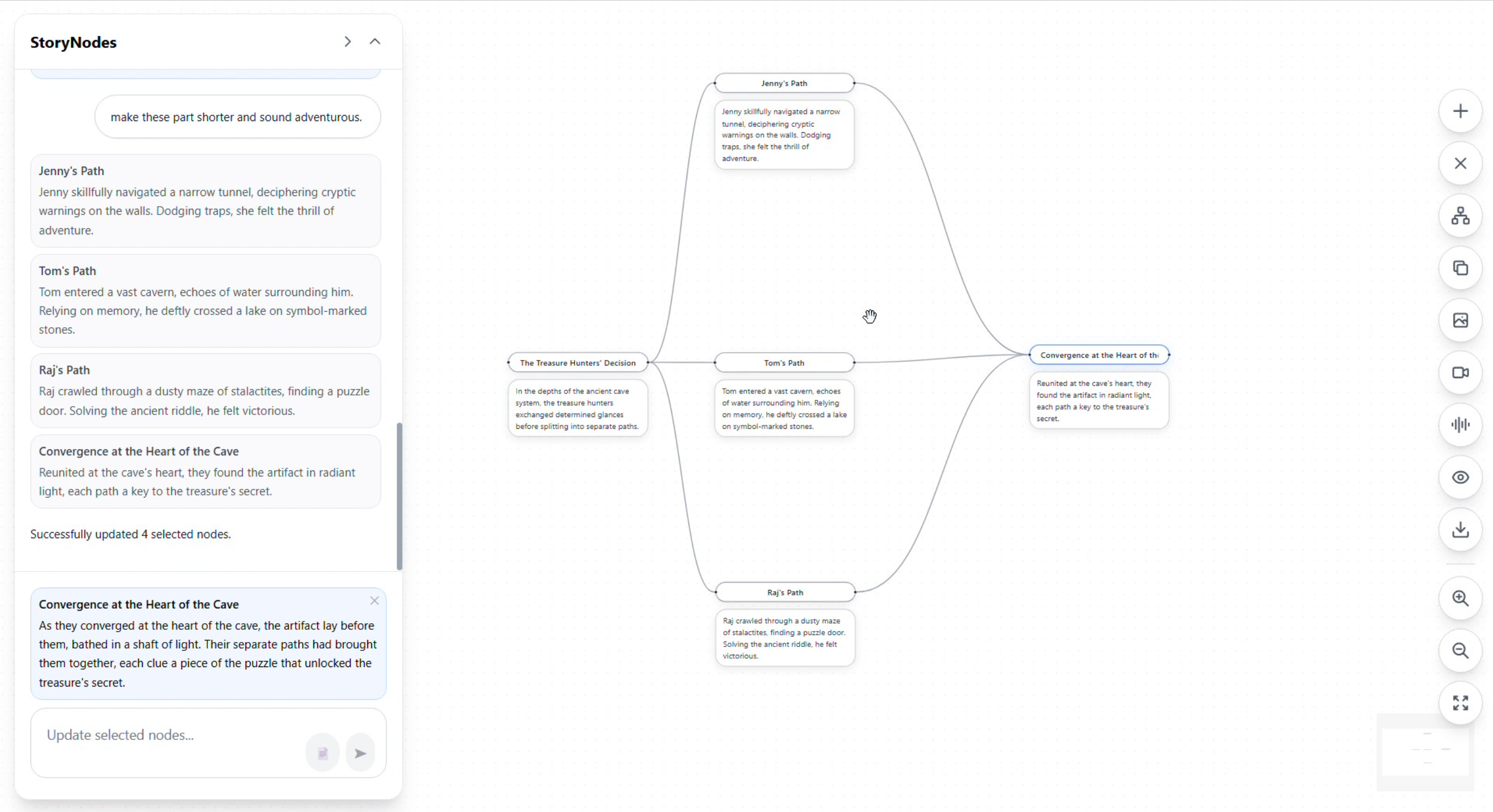}
    \caption{Updated story graph after applying the edit. Each path is rewritten in a more concise and adventurous style, demonstrating how graph-level editing modifies multiple nodes simultaneously while maintaining the branching structure}
    \label{fig:global}
\end{figure}

\begin{figure}[H]
    \centering
    \includegraphics[width=1\linewidth]{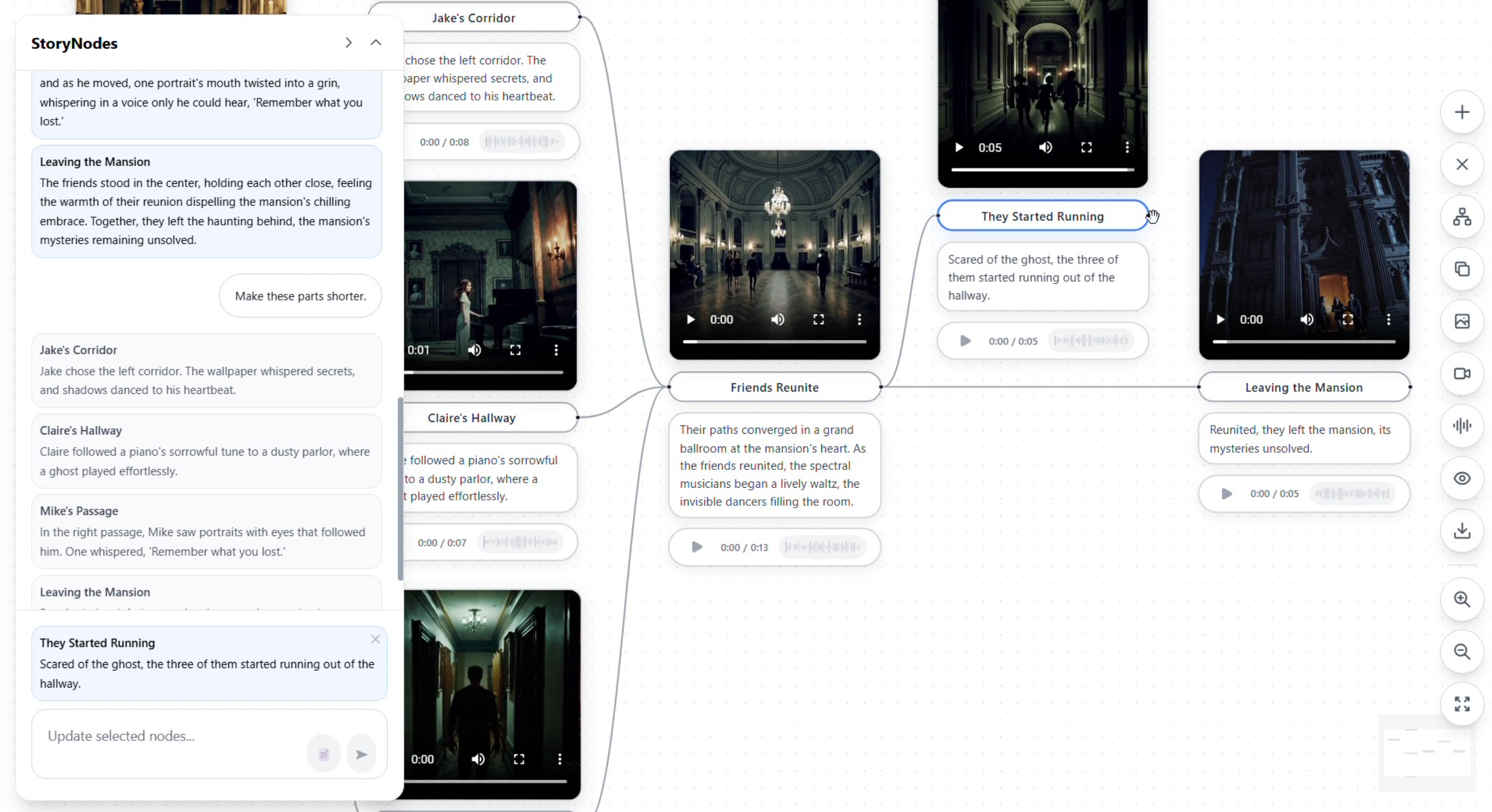}
    \caption{Node-based story extension. A new node is added introducing an additional narrative branch where the characters flee from the ghost. This demonstrates how the system allows stories to be extended by inserting new nodes, enabling users to expand the narrative structure and generate corresponding media without rewriting the entire graph.}
    \label{fig:extend}
\end{figure}

\begin{figure}[h]
    \centering
    \includegraphics[width=1\linewidth]{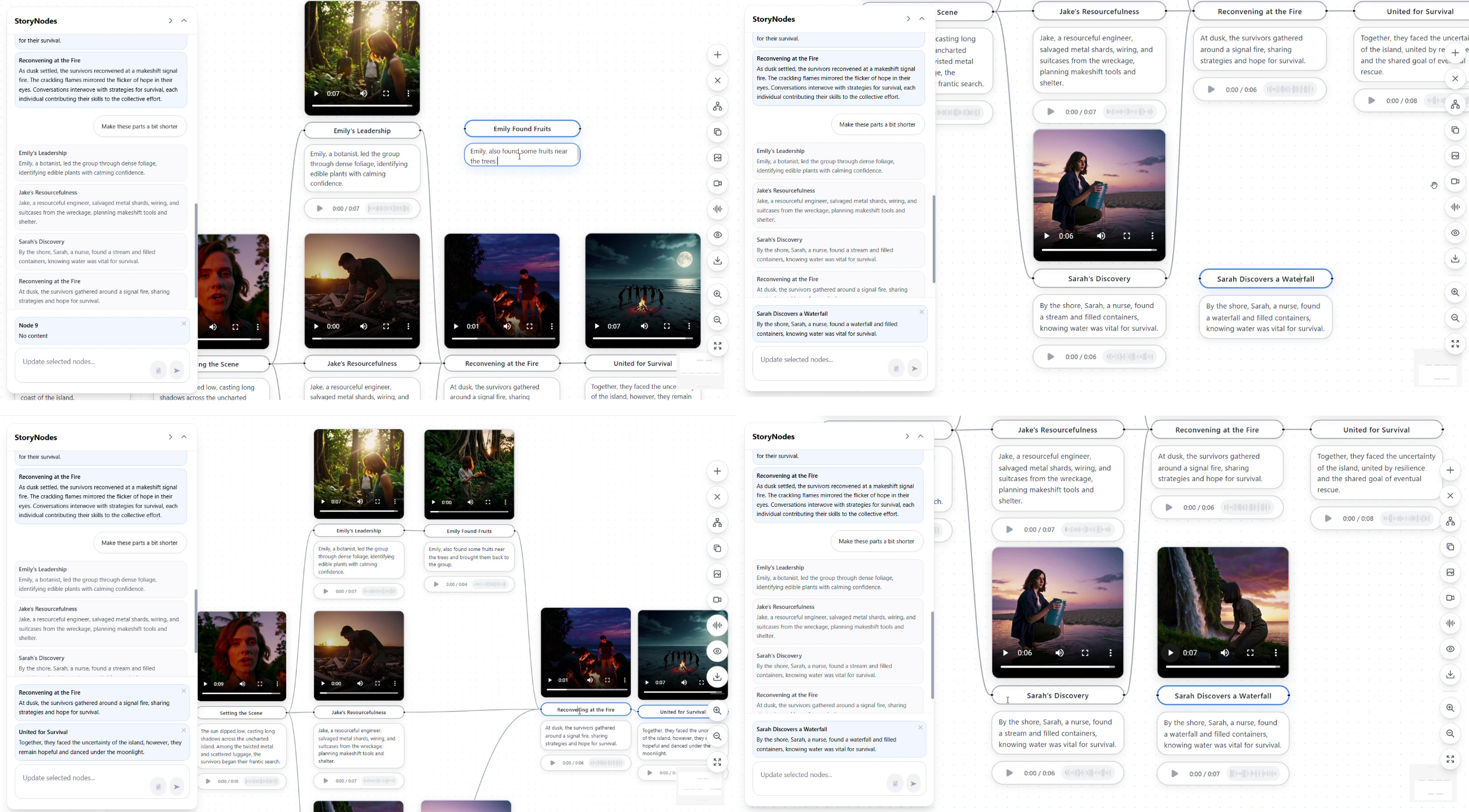}
    \caption{More example of Node Based Story Extension}
    \label{fig:extend}
\end{figure}

Preserving existing outputs while experimenting with alternatives ensures cost-effectiveness, provides creative flexibility, and allows creators to return to previous generations when needed. Branching with parallel timelines allows the system to capture events that occur simultaneously across different characters or locations. Rather than restricting the story to a single path, the graph structure enables multiple threads to unfold side by side, giving creators flexibility to model complex narratives. Such representations expand the expressive power of generative storytelling, supporting multi-perspective narratives that mirror the complexity of films, games, and interactive media. 

In addition, the system supports extending the story by adding new nodes, making it possible to introduce fresh plot points, twists, or simply build off the initial AI-generated script.

\subsection{Full video export and Compilation}\label{export}

The system supports full video export by compiling node-level media into a continuous sequence. Users can preview individual story elements, download specific assets, or export the entire narrative as a single video file with subtitles. Before exporting, the interface presents the story as a slideshow, allowing creators to review the sequence of scenes. The compilation process follows the underlying graph topology, sorting nodes according to their narrative order to ensure that the final video reflects the intended storyline. This functionality enables creators to transform interactive story graphs into polished audiovisual outputs that can be shared or archived beyond the authoring interface.

\begin{figure}[H]
    \centering
    \includegraphics[width=0.95\linewidth]{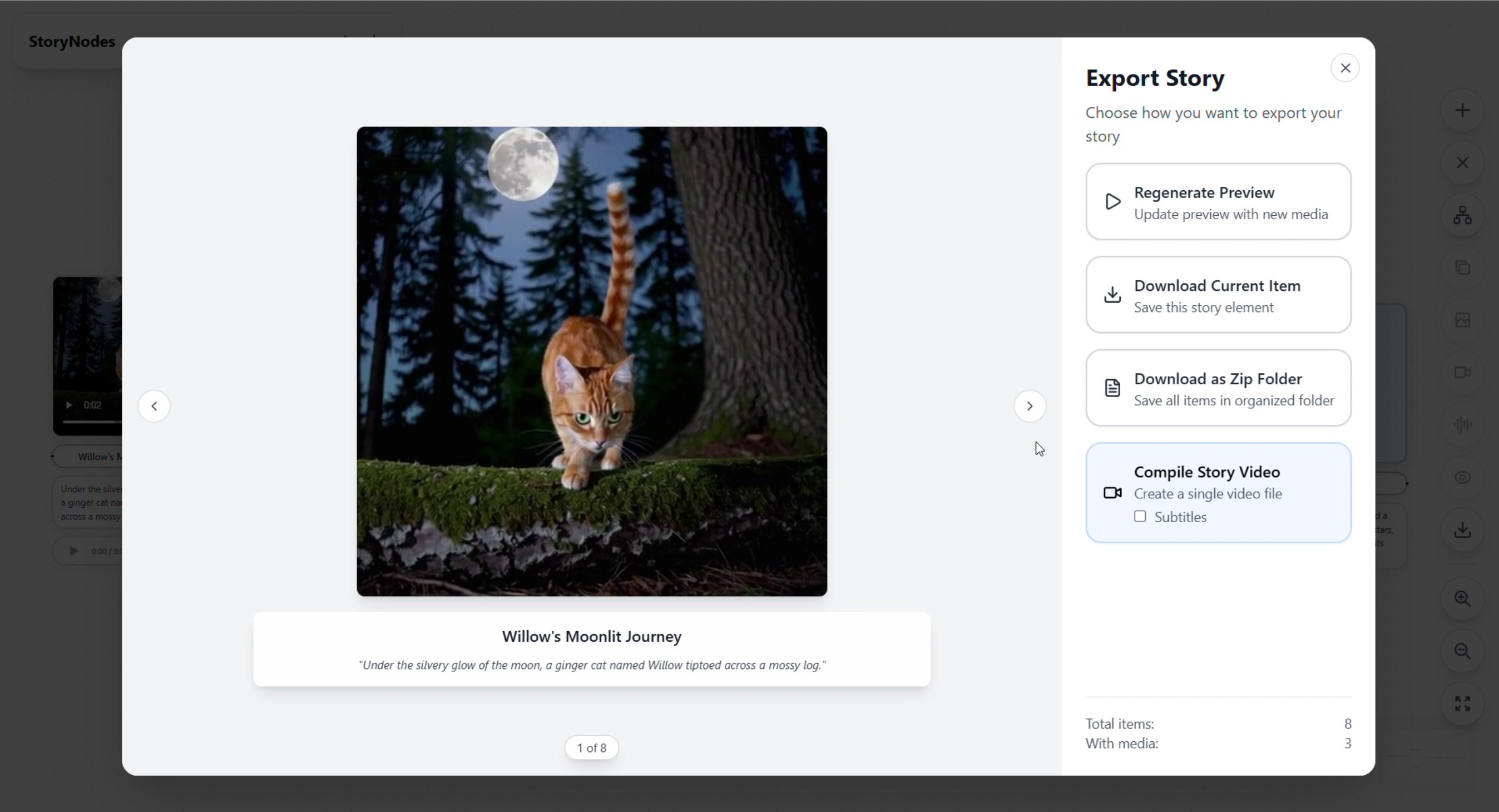}
    \caption{Export interface for full video compilation, showing options to preview, download individual items, or compile the entire story into a single video.}
    \label{fig:export}
\end{figure}

\subsection{Compute Resources and Performance} \label{compute}
All experiments in this paper were conducted using deployed generative models accessed through Azure endpoints (Sora model hosed on Azure)\cite{microsoft_azure_nodate} and OpenAI endpoints (GPT-4.1 for text and reasoning, GPT-Image-1 for images, and GPT-4o TTS for audio) \cite{openai_overview_nodate}. The orchestration code, interface, and evaluations were run locally on a standard laptop without the need for GPUs or specialized hardware. Since most computation is performed by cloud-hosted models, reproducing our results requires only API access to the same model deployments and a local Python/JavaScript environment for running the interface. Execution times were interactive in nature, with most node based image / video generation tasks completing within 10–30 seconds per API call.

\subsection{Broader Societal Impact} \label{Impact}
 By providing users with a controllable, editable, and iterative workflow, this work aims to make multimedia generative content creation easier and more accessible. Through visual, node-based human-AI interaction and generation, the system lowers the barrier for non-technical users to direct advanced models, opening opportunities for education, independent media, and community-driven storytelling. Such democratization can support more diverse voices in creative industries, allowing people without access to professional production pipelines to produce expressive multimodal works. At the same time, emphasizing controllability helps preserve human agency in the creative loop, ensuring that AI functions as a collaborator rather than a replacement. Potential negative impacts include risks of misuse for misinformation or low-quality mass content generation. These risks can be mitigated by integrating safeguards such as provenance tracking, attribution, and responsible deployment practices. In addition, protective AI techniques such as watermarking and fingerprinting are important to ensure the traceability and integrity of generated media, further reducing risks of malicious use.

%%%%%%%%%%%%%%%%%%%%%%%%%%%%%%%%%%%%%%%%%%%%%%%%%%%%%%%%%%%%

\newpage
\section*{NeurIPS Paper Checklist}

%%% END INSTRUCTIONS %%%

\begin{enumerate}

\item {\bf Claims}
    \item[] Question: Do the main claims made in the abstract and introduction accurately reflect the paper's contributions and scope?
    \item[] Answer: \answerYes{}
    \item[] Justification: The abstract and Introduction state a node-based interface for multimodal generation, a task-selection orchestration, and results on outline generation and editing workflows; these align with Methods and Experiments/Results sections.

\item {\bf Limitations}
    \item[] Question: Does the paper discuss the limitations of the work performed by the authors?
    \item[] Answer: \answerYes{}
    \item[] Justification: A dedicated Limitation and Conclusions section discusses scalability to longer narratives, cross-node consistency, and future user studies.

\item {\bf Theory assumptions and proofs}
    \item[] Question: For each theoretical result, does the paper provide the full set of assumptions and a complete (and correct) proof?
    \item[] Answer: \answerNA{}
    \item[] Justification: The paper presents a system/interface and empirical observations; it does not include theoretical results or proofs.

\item {\bf Experimental result reproducibility}
    \item[] Question: Does the paper fully disclose all the information needed to reproduce the main experimental results of the paper to the extent that it affects the main claims and/or conclusions of the paper (regardless of whether the code and data are provided or not)?
    \item[] Answer: \answerYes{}
    \item[] Justification: We provide experiment information in the appendix and method section including the roles of each LLM task in the pipeline, json file structure, the actually prompts used to test the system for linear and branching narratives, and the corresponding results. 

\item {\bf Open access to data and code}
    \item[] Question: Does the paper provide open access to the data and code, with sufficient instructions to faithfully reproduce the main experimental results, as described in supplemental material?
    \item[] Answer: \answerNo{}
    \item[] Justification: Code and assets are not currently released; the project is still in development and we intend to provide an anonymized supplement with instructions in a camera-ready version if feasible.

\item {\bf Experimental setting/details}
    \item[] Question: Does the paper specify all the training and test details (e.g., data splits, hyperparameters, how they were chosen, type of optimizer, etc.) necessary to understand the results?
    \item[] Answer: \answerYes{}
    \item[] Justification: While we do not train any new models, we specify that the experiments rely on pretrained models. The experimental settings, including the model types used (e.g., GPT-4.1, GPT-Image-1, GPT-4o TTS, Sora) and their roles in the LLM task pipeline, are described in the Methods section.

\item {\bf Experiment statistical significance}
    \item[] Question: Does the paper report error bars suitably and correctly defined or other appropriate information about the statistical significance of the experiments?
    \item[] Answer: \answerYes{}
    \item[] Justification: We report success rates across 10 trials for linear (8/10 correct) and branching narratives (10/10 correct), and include binomial confidence intervals to indicate variability. Given the small sample size, results are presented as descriptive proportions with error estimates, complemented by qualitative workflow observations.

\item {\bf Experiments compute resources}
    \item[] Question: For each experiment, does the paper provide sufficient information on the computer resources (type of compute workers, memory, time of execution) needed to reproduce the experiments?
    \item[] Answer: \answerYes{}
    \item[] Justification: The experiments were conducted by calling deployed generative models via Azure endpoints, with orchestration and interface code run locally on a standard laptop. Since the heavy computation is handled by the cloud-hosted models, reproducing results requires only access to the same APIs and modest local resources (CPU laptop for interface execution).

\item {\bf Code of ethics}
    \item[] Question: Does the research conducted in the paper conform, in every respect, with the NeurIPS Code of Ethics \url{https://neurips.cc/public/EthicsGuidelines}?
    \item[] Answer: \answerYes{}
    \item[] Justification: The work involves system design and small-scale, non-human-subjects evaluations; no sensitive data are collected and use of third-party models is disclosed. We will include an ethics statement in the final.

\item {\bf Broader impacts}
    \item[] Question: Does the paper discuss both potential positive societal impacts and negative societal impacts of the work performed?
    \item[] Answer: \answerYes{}
    \item[] Justification: Broader impacts are explicitly discussed in both the introduction, conclusion and the appendix. 

\item {\bf Safeguards}
    \item[] Question: Does the paper describe safeguards that have been put in place for responsible release of data or models that have a high risk for misuse (e.g., pretrained language models, image generators, or scraped datasets)?
    \item[] Answer: \answerYes{}
    \item[] Justification: No high-risk datasets or new pretrained models are released in this submission; the system orchestrates existing model APIs.

\item {\bf Licenses for existing assets}
    \item[] Question: Are the creators or original owners of assets (e.g., code, data, models), used in the paper, properly credited and are the license and terms of use explicitly mentioned and properly respected?
    \item[] Answer: \answerYes{}
    \item[] Justification: We properly credit all the models we used throughout the paper in the references.

\item {\bf New assets}
    \item[] Question: Are new assets introduced in the paper well documented and is the documentation provided alongside the assets?
    \item[] Answer: \answerNA{}
    \item[] Justification: No new datasets or pretrained models are released with this submission; the contribution is a novel system and a new workflow for multimodal content creation.

\item {\bf Crowdsourcing and research with human subjects}
    \item[] Question: For crowdsourcing experiments and research with human subjects, does the paper include the full text of instructions given to participants and screenshots, if applicable, as well as details about compensation (if any)? 
    \item[] Answer: \answerNA{}
    \item[] Justification: The current work does not include crowdsourcing or human-subjects studies; future user studies are listed as planned work.

\item {\bf Institutional review board (IRB) approvals or equivalent for research with human subjects}
    \item[] Question: Does the paper describe potential risks incurred by study participants, whether such risks were disclosed to the subjects, and whether Institutional Review Board (IRB) approvals (or an equivalent approval/review based on the requirements of your country or institution) were obtained?
    \item[] Answer: \answerNA{}
    \item[] Justification: No human-subjects research is reported in this submission.

\item {\bf Declaration of LLM usage}
    \item[] Question: Does the paper describe the usage of LLMs if it is an important, original, or non-standard component of the core methods in this research?
    \item[] Answer: \answerYes{}
    \item[] Justification: The Methods section details a task-selection agent and the roles of LLMs (e.g., GPT-4.1) for story generation, reasoning, diagramming, and editing within the node-based workflow.

\end{enumerate}

\end{document}